\documentclass[11pt,nofootinbib]{revtex4-1}

\usepackage{amsmath,mathrsfs}
\usepackage{graphicx,epsfig}
\usepackage{latexsym,amssymb}
\usepackage{epsf}
\usepackage{ifpdf,natbib,lineno}

\begin{document}

\title{{\bf  An   Energetically  Stable Q-ball solution  in $3+1$ Dimensions}}

\author{ M. Mohammadi} \email{physmohammadi@pgu.ac.ir} \address{Physics Department, Persian Gulf University, Bushehr 75169, Iran.}

\begin{abstract}

The paper, classically,  presents  an extended  Klein-Gordon  field system in $3+1$ dimensions  with a special   Q-ball solution.   The  Q-ball solution  is energetically stable, that is, for any arbitrary small deformation above the background of that, total energy always  increases.   The general dynamical equations,  just for this special Q-ball  solution, are reduced to the known versions  of  a complex nonlinear Klein-Gordon system,  as its  dominant dynamical equations.


\end{abstract}

\maketitle


\section{Introduction}

The  complex nonlinear Klein-Gordon (CNKG) systems with the well-known non-topological  Q-ball solutions,  have been of interest to  physicist \cite{GR,SR,Scoleman,Ros,GCM,R1,R2,R3,R4,CG1,CG2,CG3,CG4,DM1,DM2,DM3,DM4,DM5,GQ1,GQ2,GQ3,GQ4,GQ5,
GQ6,GQ7,GQ8,GQ9,GQ10,GQ11,GQ12,GQ13,GQ14,GQ15,GQ16,GQ17,GQ18,GQ19,MI,Vak3,Vak4,Vak5,Vak6,Vak7,
Vak8,Vak9,Lee3,Vak777,Vak22}.  For the first time, such   non-topological lumps  was proposed in \cite{GR} and then called  Q-balls \cite{SR}. Since the Lagrangian densities  which bear  the Q-ball solutions have the global $U(1)$ symmetry, then any  Q-ball solution has a specific charge $Q$ and a specific rest frequency $\omega_{o}$.
Q-balls are interesting for gravitational waves production and different cosmological scenarios \cite{CG1,CG2,CG3,CG4}. They are also introduced  as  dark matter candidates \cite{DM1,DM2,DM3,DM4,DM5}.
Moreover, the  gauged Q-balls have    been  of interest to many  articles \cite{GQ1,GQ2,GQ3,GQ4,GQ5,GQ6,GQ7,GQ8,GQ9,GQ10,GQ11,GQ12,GQ13,GQ14,GQ15,GQ16,GQ17,GQ18,GQ19}. In general, there is   a vast  literature on  the stationary Q-balls, for example, one can  see \cite{MI} and the references therein.

Based on these motivations, the stability of Q-balls has been intensively studied \cite{Vak3,Vak4,Vak5,Vak6,Vak7,Vak8,Vak9,Lee3,Vak777,Vak22}.   In general, the stability is the main   condition for a solitary wave solution to be a soliton. For the topological solitary wave solutions, the stability  is inherent. But, for the non-topological solitary wave solutions, there are   different criteria   for the stability depending on  purposes.  Specially, for the systems with  non-topological Q-ball solution, there are  three well-known criteria that are called the classical (Vakhitov-Kolokolov), the quantum mechanical and the fission stability criteria, respectively. The classical stability criterion is based on the examining dynamical equations  when is  linearized for the small fluctuations above the background of the solitary wave solution \cite{Vak3,Vak4,Vak5,Vak6,Vak7,Vak8,Vak9,Lee3,Vak777,Vak22,Vak1,Vak2}. A solitary wave solution which is classically stable, does not have any  growing mode and then  can not spontaneously  blowup to infinity. For the Q-ball solutions, the classical criterion leads to the condition $\frac{dQ}{d\omega_{o}}< 0$ for the stable ones \cite{Vak3,Vak4,Vak5,Vak6,Vak7,Vak8,Vak9,Lee3,Vak777,Vak22}.   The quantum mechanical criterion  for a typical Q-ball solution is based on the comparison  between the rest energy of  that $E_{o}$ and the rest energy of the lightest possible scalar particle quanta. A Q-ball solution which is quantum mechanically stable,  can not decay to many free scalar particle quanta. In general, if the ratio between the rest
energy and the charge is less than $\omega_{+}$ (i.e. $E_{o}/Q<\omega_{+}$),  a quantum mechanically stable Q-ball  exists \cite{Vak6,Lee3},  where $\omega_{+}$  ($\omega_{-}$) is the maximum (minimum)  on the range  of the possible rest frequencies $\omega_{-}\leqslant |\omega_{o}|\leqslant \omega_{+}$, which yield Q-ball solutions. A Q-ball may decay into two or more smaller Q-balls, if such a Q-ball does not fulfill  the fission stability condition. It was shown that the condition for the fission stability  is identical to the condition of the classical stability  \cite{Vak6}. In other words, a Q-ball solution which is classically stable would be  stable against fission too.

There is another  stability criterion, called the energetically stability criterion \cite{MM1}. If for  a solitary wave solution, any arbitrary (permissible or impermissible) deformation above the background of that leads to an increase in  the total energy, it would be indeed energetically a stable solution. In other words, an energetically stable solitary wave solution has the minimum rest energy among the other (close) solutions. In this case, unlike the Vakhitov-Kolokolov  criterion \cite{Vak3,Vak4,Vak5,Vak6,Vak7,Vak8,Vak9,Lee3,Vak777,Vak22,Vak1,Vak2}, we examine the energy density functional for the small variations   instead of dynamical equations \cite{MM1,MM2,MM3,Derrick}. In general, none of the Q-ball solutions are energetically stable objects \cite{MM1}.

In this paper in line   with \cite{MM1,MM2}, we are going to introduce an extended  KG system\footnote{Briefly, for a set of real scalar fields $\phi_{j}$ ($j=1,2,\cdots,N$), the extended KG systems have Lagrangian densities which are not linear in the kinetic scalars ${\cal S}_{ij}=\partial_{\mu}\phi_{i}\partial^{\mu}\phi_{j}$ \cite{MM2,MM3}. For example, in Refs.~\cite{R2,MM1,MM2,MM3,Vash}, the extended KG systems are used.}  in $3+1$ dimensions which leads to a special energetically stable Q-ball solution.
We show that the general  dynamical equations, just for this  special Q-ball solution,  are reduced to the known versions  of a special CNKG system, as its dominant dynamical equations.
In \cite{MM1,MM2}, there were introduced  extended KG systems which lead to  special Q-ball solutions in $1+1$ dimensions. The main idea was   to add a proper  additional term $F$ to the original standard CNKG Lagrangian density, which   guarantees the uniqueness and energetically stability of one of its Q-ball solutions. However, to bring this idea to life in $3+1$ dimensions, unlike the pervious works in $1+1$ dimensions \cite{MM1,MM2},  we have to reintroduce  the additional term $F$ using   three new scalar catalyzer fields $\psi_{1}$, $\psi_{2}$ and $\psi_{3}$, whose roles  in dominant dynamical equations and other observable of the special Q-ball solution are ineffective.
In fact, these  catalyzer fields $\psi_{j}$ ($j=1,2,3$) must be included  in the additional term $F$  to play the expected roles properly. This paper is, especially,  in line with \cite{MM1}, hence the other complementary discussions are the same as those  sufficiently presented   in \cite{MM1}.

The organization of this paper is as follows: In the next section,  for  the CNKG systems  we will review   the basic equations and consider general properties of the related Q-ball  solutions, especially a CNKG system with Gaussian Q-ball solution will be introduced in detail.  In section III,  an extended KG system with a special Q-ball solution will be introduced in $3+1$ dimensions. In section IV, the energetically  stability of the special Q-ball solution will be considered in general.  The last section is devoted to  summary and conclusions.

\section{ Basic properties of  the CNKG systems with the Q-ball solutions}\label{sec2}

For a single complex scalar field $\phi$, the relativistic $U(1)$ (or the CNKG) Lagrangian densities   with the Q-ball solutions are  defined as follows:
 \begin{equation} \label{lag}
{\cal L}_{o}= \partial_\mu \phi^*
\partial^\mu \phi -V(|\phi |) ,
 \end{equation}
in which  $V(|\phi |)$, the field potential, is a self-interaction term  which depends only on the modulus of the scalar field. By varying this action with respect to $\phi^{*}$, one obtains
the field equation
\begin{equation} \label{eq}
 \Box \phi =\frac{\partial^2\phi}{\partial
  t^2}- \nabla^{2} \phi=-\frac{\partial V}{\partial
  \phi^*}=-\frac{1}{2}\frac{d V}{d|\phi|}\frac{\phi}{|\phi|},
  \end{equation}
which is the same complex non-linear  Klein-Gordon equation in $3+1$ dimensions. Note that, through the paper,   we take the speed of light equals to one.  To simplify  Eq.~(\ref{eq}), we can change variables to the polar fields $R(x^{\mu})$ and $\theta(x^{\mu})$ as defined by
\begin{equation} \label{polar}
 \phi(x,y,z,t)= R(x,y,z,t)\exp[i\theta(x,y,z,t)].
\end{equation}
In terms of   polar fields, equivalently,  the Lagrangian density (\ref{lag}) and the related dynamical field equation (\ref{eq}), respectively,   turn  into
\begin{equation} \label{Lag2}
{\cal L}_{o}=(\partial^\mu R\partial_\mu R) +R^{2}(\partial^\mu\theta\partial_\mu\theta)-V(R),
\end{equation}
and
\begin{eqnarray} \label{e25}
&&\Box R-R(\partial^\mu\theta\partial_\mu\theta)=-\frac{1}{2}\frac{dV}{dR}, \\&&\label{e252}
\partial_{\mu}(R^2\partial^{\mu}\theta)=0,
\end{eqnarray}
respectively. The related Hamiltonian (energy) density  is obtained via the Noether's theorem:
\begin{eqnarray} \label{TE}
&&\varepsilon_{o}=\dot{\phi}\dot{\phi}^{*}+\nabla\phi \cdot \nabla\phi^{*}+V(|\phi |)=\dot{R}^2+(\nabla R)^2+R^2[\dot{\theta}^2+(\nabla\theta)^2]+V(R),
\end{eqnarray}
where dot  denotes differentiation with respect to $t$.

In general, the spherically symmetric Q-ball  solutions are introduced as follows:
\begin{equation} \label{So1}
R(x,y,z,t)=R(r)=R(\sqrt{x^2+y^2+z^2}),\quad  \quad\theta(x,y,z,t)=\omega_{o}t,
\end{equation}
in which $R(r)$ should  be  a localized function. For   ansatz (\ref{So1}), Eq.~(\ref{e252}) is satisfied automatically and Eq.~(\ref{e25}) would be reduced  to
 \begin{equation} \label{Re}
 \dfrac{1}{r^2}\dfrac{d}{dr}(r^2\dfrac{dR}{dr})=\frac{1}{2}\frac{dV}{dR}-\omega_{o}^{2}R.
 \end{equation}
 Depending on  different values of   $\omega_{o}$, different solutions  for $R(r)$ can be obtained. Accordingly,  there are infinite spherically symmetric Q-ball  solutions which characterized   by  different rest frequencies  $\omega_{-}<|\omega_{o}|<\omega_{+}$.
A moving Q-ball solution can be obtained easily by a relativistic boost. For example, for a Q-ball  solution with rest frequency $\omega_{o}$, which moves in the $x$-direction with a constant velocity $\textbf{v}=v\widehat{i}$, we have:
\begin{equation} \label{So}
R(x,y,z,t)=R(\sqrt{\gamma^2(x-vt)^2+y^2+z^2}),\quad  \quad\theta(x,y,z,t)=k_{\mu}x^{\mu},
\end{equation}
in which  $\gamma=1/\sqrt{1-v^2}$, and  $k^{\mu}\equiv(\omega,\textbf{k})=(\omega,k,0,0)$ is a $3+1$ vector, provided $\textbf{k}=k\widehat{i}= {\omega}\textbf{v}$ and $\omega=\gamma\omega_{o}$.

For simplicity,  to obtain  different   Q-ball solutions with the Gaussian modules,  one can use the following field potential:
\begin{equation} \label{fp}
V(R)= R^2\left[\lambda^{-2}+l^{-2}-l^{-2}\ln{(a^{n-1}R^2)}\right],
\end{equation}
in which, $\lambda$, $l$ and $a$ are dimensional parameters and $n$
stands for the number of spatial dimensions. This model (\ref{fp}), was proposed for the first time in \cite{Ros} and thoroughly examined in \cite{GCM}.
By solving  equation (\ref{Re}),   the variety of    Q-ball solutions as a function of  $\omega_{o}$ can be obtained:
\begin{equation} \label{f}
R(r)=A(\omega_{o})\exp{\left(-\frac{r^{2}}{2l^2}\right)},
\end{equation}
where $0\leqslant |\omega_{o}|\leqslant \infty$, and
\begin{equation} \label{A}
A(\omega_{o})=a^{^{(\frac{1-n}{2})}}\exp{\left(\frac{n+(l/\lambda)^2-(\omega_{o}l)^2}{2}\right)}.
\end{equation}
 The total energy of a non-moving  Q-ball solution can be obtained and equated to the rest energy  of that   as
\begin{eqnarray} \label{fe}
&&E_{o}(\omega_{o})=m_{o}\equiv \int T^{00} d^{3}\textbf{x}=\int \left[(\nabla R)^2 +R^2(\dot{\theta}^2)+V(R)  \right]d^{3}\textbf{x}\nonumber \\&&
=\int_{0}^{\infty} \left[(\frac{dR}{dr})^{2}+\omega_{o}^{2}R^{2}+V(R)\right]4\pi r^{2}dr =(C/l)[(l\omega_{o})^2+\frac{1}{2}]\exp{\left(-(l\omega_{o})^2\right)},
\end{eqnarray}
where $C=2\sqrt{\pi}(l\sqrt{\pi}/a)^{n-1}\exp{[n+(l/\lambda)^2]}$.


The Lagrangian density (\ref{lag}) is $U(1)$ invariant like electromagnetic theory  and this yields to the conservation of the electrical  charge. So, according to the Noether theorem, we can introduce a conserved electrical   current density as
\begin{equation} \label{cur}
j^\mu\equiv i (\phi\partial^\mu \phi^* -\phi^* \partial^\mu \phi)=2 (R^{2}\partial^\mu \theta),
\end{equation}
where  $\partial_{\mu} j^{\mu}=0$. Therefore, the corresponding conserved charge would be
\begin{equation} \label{Bar}
Q(\omega_{o})=\int j^0 d^{3}\textbf{x}=2\omega_{o}\int R^{2}d^{3}\textbf{x}=
Cl\omega_{o}\exp{\left(-(l\omega_{o})^2\right)}.
\end{equation}
 It is notable that  both  positive and negative signs of $|\omega_{o}|$  (i.e. $\omega_{o}=\pm |\omega_{o}|$) lead to the same  solution for the differential equation (\ref{Re}). They have the same rest mass (energy) but different electrical   charges (positive and negative). It is easy to show that for the solutions with $\omega_{o}>0$ ($\omega_{o}<0$),  the electrical  charge is positive (negative).

Now, we can study the stability of the  Gaussian Q-balls (\ref{f})   based on the different known stability criteria. Since $\omega_{+}=\infty$ and condition $E_{o}/Q<\omega_{+}$ is fulfilled for all Q-balls (\ref{f}),  thus all of them are quantum mechanically stable. The condition $\frac{dQ}{d\omega_{o}}< 0$ leads to inequality $\omega_{o}^2>1/2l^2$  (see \cite{GCM}) for the Q-balls (\ref{f}) which are classically stable and stable against fission too. In the next sections, we will show how adding  a proper term to the Lagrangian density (\ref{lag}) yields    a special   energetically stable   Q-ball solution as well.

\section{An  extended  KG system  with a special  Q-ball solution}\label{sec4}

Similar to the remarks  made  at the   beginning of the section 4 (3) of the Ref.~\cite{MM1} (\cite{MM2}), we are going to consider a new Lagrangian density as follows:
\begin{equation} \label{LN}
 {\cal L}= {\cal L}_{o}+F=\left[\partial^\mu R\partial_\mu R +R^{2}(\partial^\mu\theta\partial_\mu\theta)-V(R)\right]+ F,
 \end{equation}
where $F$ is considered to be a proper additional term  whose responsibility   is to guarantee the  uniqueness   and the energetically stability of a  special Q-ball solution; meaning that, it should behave  as a stability catalyzer just for a special  Q-ball solution.   Moreover, $F$ and all of its  derivatives  should be zero just for the special Q-ball solution. Suppose that the  special Q-ball  solution is as follows:
\begin{equation} \label{f2}
\phi_{s}(r,t)=R_{s}(r)e^{i\theta_{s}}=\exp{\left( \frac{-r^2}{2}\right)}\exp{(i\omega_{s}t)},
\end{equation}
where $\omega_{s}=\sqrt{2}$. In fact, it is one of the introduced Q-ball solutions (\ref{f}) for which $l=\lambda=1$, $a=e^{1}$ and $\omega_{o}=\omega_{s}=\sqrt{2} t$; hence  $V(R)=-2R^2\ln R$, $R_{s}(r)=\exp{\left( \frac{-r^2}{2}\right)}$. Since $\omega_{s}^2>1/2$, it is  a classical stable Q-ball solution obviously.

In fact, we are going to build a new classical relativistic field system in such a way that the general dynamical equations belong to Lagrangian density (\ref{LN}) are reduced to the same standard  versions (\ref{e25}) and (\ref{e252}) just for the special  Q-ball solution (\ref{f2}), as its  dominant dynamical equations. Moreover, as we indicated before, this  special Q-ball solution (\ref{f2}) should be an energetically stable object. To meet  these requirements, we can propose a proper additional term in the  following form:
\begin{equation} \label{kk}
F=B\sum_{i=1}^{12}{\cal K}_{i}^3,
 \end{equation}
in which $B$ is considered to be a large number. Functionals ${\cal K}_{i}$'s are defined as follows:
\begin{eqnarray} \label{jj}
&&{\cal K}_{1}=R^2\mathbb{S}_{2},\quad\quad~~~ {\cal K}_{2}=R^2h_{2}^2\mathbb{S}_{2}+\mathbb{S}_{1},\quad\quad~~~ {\cal K}_{3}=R^2h_{3}^2\mathbb{S}_{2}+\mathbb{S}_{1}+2Rh_{3}\mathbb{S}_{3},  \nonumber \\&&\label{jj1}
 {\cal K}_{4}=R^2[h_{4}^2\mathbb{S}_{2}+\mathbb{S}_{4}],\quad\quad~~~ {\cal K}_{5}=R^2[h_{5}^2\mathbb{S}_{2}+\mathbb{S}_{5}],\quad\quad~~~ {\cal K}_{6}=R^2[h_{6}^2\mathbb{S}_{2}+\mathbb{S}_{6}], \nonumber \\&&\label{jjf}
{\cal K}_{7}=R^2[h_{7}^2\mathbb{S}_{2}+\mathbb{S}_{4}+\mathbb{S}_{5}+2\mathbb{S}_{7}],\quad\quad~~~ {\cal K}_{8}=R^2[h_{8}^2\mathbb{S}_{2}+\mathbb{S}_{4}+\mathbb{S}_{6}+2\mathbb{S}_{8}],\nonumber  \\&&\label{jj6}
{\cal K}_{9}=R^2[h_{9}^2\mathbb{S}_{2}+\mathbb{S}_{5}+\mathbb{S}_{6}+2\mathbb{S}_{9}], \quad\quad~~~ {\cal K}_{10}= R^2h_{10}^2 \mathbb{S}_{2}+\mathbb{S}_{1}+R^2\mathbb{S}_{4}+2R\mathbb{S}_{10}, \nonumber\\&&\label{jj9}
{\cal K}_{11}= R^2h_{11}^2 \mathbb{S}_{2}+\mathbb{S}_{1}+R^2\mathbb{S}_{5}+2R\mathbb{S}_{11},~~~
{\cal K}_{12}= R^2h_{12}^2 \mathbb{S}_{2}+\mathbb{S}_{1}+R^2\mathbb{S}_{6}+2R\mathbb{S}_{12},
\end{eqnarray}
where
\begin{eqnarray} \label{sd}
&&\mathbb{S}_{1}=\partial_{\mu}R\partial^{\mu}R-2R^2\ln R,\quad~~~ \mathbb{S}_{2}=\partial_{\mu}\theta\partial^{\mu}\theta-2,\quad~~~ \mathbb{S}_{3}=\partial_{\mu}R\partial^{\mu}\theta, \nonumber \\&&\label{ll1}
\mathbb{S}_{4}=\partial_{\mu}\psi_{1}\partial^{\mu}\psi_{1}+R^2-2\psi_{1}^2(\ln R+1),\quad~~~\mathbb{S}_{5}=\partial_{\mu}\psi_{2}\partial^{\mu}\psi_{2}+R^2-2\psi_{2}^2 (\ln R+1), \nonumber \\&&\label{ll2}
\mathbb{S}_{6}=\partial_{\mu}\psi_{3}\partial^{\mu}\psi_{3}+R^2-2\psi_{3}^2 (\ln R+1),\quad~~~\mathbb{S}_{7}=\partial_{\mu}\psi_{1}\partial^{\mu}\psi_{2}-2\psi_{1}\psi_{2}(\ln R+1),\nonumber\\&&\label{ll3}
\mathbb{S}_{8}=\partial_{\mu}\psi_{1}\partial^{\mu}\psi_{3}-2\psi_{1}\psi_{3}(\ln R+1),\quad~~~
\mathbb{S}_{9}=\partial_{\mu}\psi_{2}\partial^{\mu}\psi_{3}-2\psi_{2}\psi_{3}(\ln R+1),\nonumber\\&&\label{ll4}
\mathbb{S}_{10}=\partial_{\mu}R\partial^{\mu}\psi_{1}-R\psi_{1}(2\ln R+1), \quad~~~\mathbb{S}_{11}=\partial_{\mu}R\partial^{\mu}\psi_{2}-R\psi_{2}(2\ln R+1),\nonumber\\&&\label{ll5}
\mathbb{S}_{12}=\partial_{\mu}R\partial^{\mu}\psi_{3}-R\psi_{3}(2\ln R+1).
\end{eqnarray}
and
\begin{eqnarray} \label{hh}
&&h_{2}=h_{3}=\frac{1}{2}[\ln R-1],\quad~~ h_{4}=\frac{1}{2}[\psi_{1}^2(1+\ln R)-\frac{1}{2}R^2-1],\quad~~ h_{5}=\frac{1}{2}[\psi_{2}^2(1+\ln R)-\frac{1}{2}R^2-1],  \nonumber \\&&\label{jj1}
h_{6}=\frac{1}{2}[\psi_{3}^2(1+\ln R)-\frac{1}{2}R^2-1], \quad ~~~ h_{7}=\frac{1}{2}[(\psi_{1}+\psi_{2})^2(1+\ln R)-R^2-1], \nonumber \\&&\label{jjf}
h_{8}=\frac{1}{2}[(\psi_{1}+\psi_{3})^2(1+\ln R)-R^2-1],\quad\quad~~~ h_{9}=\frac{1}{2}[(\psi_{2}+\psi_{3})^2(1+\ln R)-R^2-1],\nonumber  \\&&\label{jj6}
h_{10}=\frac{1}{2}[(1+\psi_{1})^2\ln R+\psi_{1}^2+\psi_{1}-\frac{1}{2}R^2-1],\quad~~ h_{11}=\frac{1}{2}[(1+\psi_{2})^2\ln R+\psi_{2}^2+\psi_{2}-\frac{1}{2}R^2-1], \nonumber\\&&\label{jj9}
h_{12}=\frac{1}{2}[(1+\psi_{3})^2\ln R+\psi_{3}^2+\psi_{3}-\frac{1}{2}R^2-1],
\end{eqnarray}
in which  $\psi_{1}$, $\psi_{2}$ and $\psi_{3}$ are three new  scalar fields which can be called the catalyzer fields. We build this new system (\ref{LN}) deliberately in such a way that there is just  a unique non-trivial   common  solution  for  twelve independent  conditions $\mathbb{S}_{i}=0$ ($i=1,2,\cdots,12$) as follows:
\begin{equation} \label{ss}
R=\exp{\left( \frac{-r^2}{2}\right)},\quad \theta=\omega_{s}t,\quad \psi_{j}=x^{j}\exp{\left( \frac{-r^2}{2}\right)}\quad (j=1,2,3),
\end{equation}
where $x^{1}=x$, $x^{2}=y$ and $x^{3}=z$. Note that, the form  of $R$  and $\theta$ in  (\ref{ss}) are the same components of the  proposed  special Q-ball solution (\ref{f2}).    Twelve conditions $\mathbb{S}_{i}=0$ ($i=1,2,\cdots,12$) can be considered as twelve independent  PDEs for  five scalar fields $R$, $\theta$, $\psi_{j}$ ($j=1,2,3$); therefore, except  (\ref{ss}),  there should be no common solution as a rule. Moreover,  since twelve functionals ${\cal K}_{i}$'s ($i=1,2,\cdots,12$) are introduced  as twelve independent linear combinations of $\mathbb{S}_{i}$'s, therefore, both twelve independent  conditions  $\mathbb{S}_{i}$'s$=0$ and  ${\cal K}_{i}$'s$=0$ are equivalent.

Similar to  \cite{MM1,MM2}, if we do not use three catalyzer  fields $\psi_{j}$ ($j=1,2,3$), there are just three scalar functionals $\mathbb{S}_{1}$, $\mathbb{S}_{2}$ and $\mathbb{S}_{3}$ for which the conditions $\mathbb{S}_{i}$'s$=0$ ($i=1,2,3$)  lead to infinite independent common solutions such as:
\begin{equation} \label{sgh}
R=\exp{\left( \frac{-(r+\xi)^2}{2}\right)},\quad \theta=\omega_{s}t,
\end{equation}
where $\xi$ is any arbitrary real number. Note that, the case $\xi=0$ is the same proposed   special solution  (\ref{f2}). In fact, for any  static module function  $R=R(x,y,z)$  along with  $\theta=\omega_{s}t$,  conditions $\mathbb{S}_{2}=0$ and $\mathbb{S}_{3}=0$ are satisfied automatically. Hence the condition $\mathbb{S}_{1}=0$ is reduced to
\begin{equation} \label{fi}
(\nabla R)^2+2R^2\ln R=0,
\end{equation}
which  is a static non-linear PDE  in $3+1$ dimensions with infinite solutions such as $R=\exp{(-(r+\xi)^2/2)}$.
Therefore, since three  conditions $\mathbb{S}_{i}$'s$=0$ ($i=1,2,3$) in $3+1$ dimensions do not yield a unique  common solution, we have to consider a more complected system (\ref{LN}) with three new catalyzer  fields $\psi_{j}$ ($j=1,2,3$). Now,  twelve  conditions $\mathbb{S}_{i}$'s$=0$ ($i=1,2,\cdots,12$) exist for five fields $R$, $\theta$ and $\psi_{j}$ ($j=1,2,3$) in such a way that the module field $R$  contributes  in  ten  new conditions $\mathbb{S}_{i}$'s$=0$ ($i=3,4,\cdots,12$)  and leads  to a unique  common solution (\ref{ss}) for $\mathbb{S}_{i}$'s$=0$ ($i=1,2,\cdots,12$) simultaneously.


Using the  Euler-Lagrange equations for the new Lagrangian density (\ref{LN}), one can obtain the related dynamical equations  easily:
\begin{eqnarray} \label{jkt}
&&\left\{\Box R-R(\partial^\mu\theta\partial_\mu\theta)+\frac{1}{2}\frac{dV}{dR}\right\}+\dfrac{3B}{2}\sum_{i=1}^{12} \left[2{\cal K}_{i}(\partial_{\mu}{\cal K}_{i})   \frac{\partial{\cal K}_{i}}{\partial(\partial_{\mu}R)}    +   {\cal K}_{i}^2\partial_{\mu}\left(\frac{\partial{\cal K}_{i}}{\partial(\partial_{\mu}R)}\right)    -    {\cal K}_{i}^2\frac{\partial{\cal K}_{i}}{\partial R}   \right]=0,~~~~~~~\\&&\label{jkt1}
\left\{\partial_{\mu}(R^2\partial^{\mu}\theta)\right\}+\dfrac{3B}{2}\sum_{i=1}^{12} \left[2{\cal K}_{i}(\partial_{\mu}{\cal K}_{i})   \frac{\partial{\cal K}_{i}}{\partial(\partial_{\mu}\theta)}    +   {\cal K}_{i}^2\partial_{\mu}\left(\frac{\partial{\cal K}_{i}}{\partial(\partial_{\mu}\theta)}\right)       \right]=0.\\&&\label{jkt2}
\sum_{i=1}^{12} \left[2{\cal K}_{i}(\partial_{\mu}{\cal K}_{i})   \frac{\partial{\cal K}_{i}}{\partial(\partial_{\mu}\psi_{j})}    +   {\cal K}_{i}^2\partial_{\mu}\left(\frac{\partial{\cal K}_{i}}{\partial(\partial_{\mu}\psi_{j})}\right)    -    {\cal K}_{i}^2\frac{\partial{\cal K}_{i}}{\partial \psi_{j}}   \right]=0, \quad (j=1,2,3).
\end{eqnarray}
In general, these equations, (\ref{jkt})-(\ref{jkt2}), are very  complicated, but there is  a single special solution (\ref{ss}) for which all terms which contain ${\cal K}_{i}$'s  and ${\cal K}_{i}^2$'s (i.e. the terms which are in the brackets) would be zero simultaneously.  Therefore, for the special solution  (\ref{ss}), Eq.~(\ref{jkt2}) satisfies automatically and Eqs.~(\ref{jkt}) and (\ref{jkt1}) are reduced  to
\begin{eqnarray} \label{dj}
&&\left\{\Box R-R(\partial^\mu\theta\partial_\mu\theta)+\frac{1}{2}\frac{dV}{dR}\right\}   =0,\\&&\label{dj1}
\left\{\partial_{\mu}(R^2\partial^{\mu}\theta)\right\}=0,
\end{eqnarray}
 which  are the same as standard CNKG equations (\ref{e25}) and (\ref{e252}) respectively. It is obvious that the set of the module part $R$ and the phase part $\theta$ of (\ref{ss}) satisfy    equations (\ref{dj}) and  (\ref{dj1}) too, as we expected. In other words,  the  complicated dynamical equations (\ref{jkt})-(\ref{jkt2})  are reduced  to the same simple original dynamical equations (\ref{e25}) and (\ref{e252}) just for a  special  solution (\ref{ss}), whose  module and phase parts build   a special Q-ball solution (\ref{f2}); meaning that,  the standard Eqs.~(\ref{e25}) and (\ref{e252}) are now the dominant dynamical equations just for a  special  Q-ball  solution (\ref{f2}).  The other Q-ball solutions (\ref{f}) of the original Lagrangian density (\ref{lag})  are no longer  the solutions of the new  system (\ref{LN}). The solution (\ref{ss}) should  be called a  special Q-ball solution exactly, along with three catalyzer fields  $\psi_{1}$, $\psi_{2}$ and $\psi_{3}$, but we can only call it ``\emph{the special (Q-ball) solution}" in the rest of the article for simplicity.
Note that, the additional term $F$  in the new system (\ref{LN})   guarantees  the uniqueness  of the special  solution (\ref{ss}); meaning that, there is just a unique  special solution (\ref{ss}) for which all ${\cal K}_{i}$'s ($i=1,2,\cdots,12$) are zero simultaneously, or just for the  special solution (\ref{ss}) the dominant dynamical equations are the same standard CNKG versions (\ref{e25}) and (\ref{e252}). Moreover, in the next section we will show that $F$ guarantees the energetically stability of the special solution (\ref{ss}) as well.

It should be note that, since the Lagrangian density (\ref{LN}) is essentially Poincar\'{e} invariant, instead of the special  solution (\ref{ss}), any arbitrary  spatially  rotated version  can be used equivalently. For example, instead of (\ref{ss})  we can perform  any  rotation about $z$-axis:
\begin{eqnarray} \label{ssr1}
R=\exp{\left( \frac{-r^2}{2}\right)},\quad \theta=\omega_{s}t,\quad \psi_{1}=(\cos(\alpha)x+\sin(\alpha)y)\exp{\left( \frac{-r^2}{2}\right)},\nonumber\\ \label{ssr2}
\psi_{2}=(-\sin(\alpha)x+\cos(\alpha)y)\exp{\left( \frac{-r^2}{2}\right)}, \quad \psi_{3}=z\exp{\left( \frac{-r^2}{2}\right)},
\end{eqnarray}
 where $\alpha$ is an arbitrary angle. Moreover, using a relativistic boost, one can obtain easily the moving version of the  special  solution (\ref{ss}). For example, if it moves in the $x$-direction, we have
\begin{eqnarray} \label{cvb}
&& R=\exp\left({\frac{1}{2}[\gamma^2 (x-vt)^2+y^2+z^2}]\right), \quad \theta=k_{\mu}x^{\mu}, \quad \psi_{1}=\gamma (x-vt)\exp\left({\frac{1}{2}[\gamma^2 (x-vt)^2+y^2+z^2}]\right),\nonumber\\ \label{cv2}&&
\psi_{2}=y\exp\left({\frac{1}{2}[\gamma^2 (x-vt)^2+y^2+z^2}]\right),\quad \psi_{3}=z\exp\left({\frac{1}{2}[\gamma^2 (x-vt)^2+y^2+z^2}]\right)
\end{eqnarray}
where $k^{\mu}\equiv (\gamma \omega_{s},\gamma \omega_{s}v,0,0)$.

\section{energetically stability of the special  solution}\label{sec5}

The energy-density  of  the new extended Lagrangian-density (\ref{LN}),  is
\begin{eqnarray} \label{nnmn}
&&\varepsilon=\frac{\partial {\cal L}}{\partial\dot{R}}\dot{R}+\frac{\partial {\cal L}}{\partial\dot{\theta}}\dot{\theta}+\sum_{j=1}^{3}\frac{\partial {\cal L}}{\partial\dot{\psi_{j}}}\dot{\psi_{j}}-{\cal L}=\varepsilon_{o}+\sum_{i=1}^{12}\varepsilon_{i}=\varepsilon_{o}+B\sum_{i=1}^{12}{\cal K}_{i}^{2}\left[3C_{i}
-{\cal K}_{i}\right],
\end{eqnarray}
which is  divided  into thirteen   distinct  parts, in which
\begin{equation}\label{cof}
C_{i}=\dfrac{\partial{\cal K}_{i}}{\partial \dot{\theta}}\dot{\theta}+\dfrac{\partial{\cal K}_{i}}{\partial \dot{R}}\dot{R}+\sum_{j=1}^{3}\dfrac{\partial{\cal K}_{i}}{\partial \dot{\psi_{j}}}\dot{\psi_{j}}=
\begin{cases}
 2R^2\dot{\theta}^{2} & \text{i=1}
\\
2[\dot{R}^{2}+R^2\dot{\theta}^2h_{2}^2] & \text{i=2}
\\2[\dot{R}+R\dot{\theta}h_{3}]^2
 & \text{i=3}.
 \\2 R^2[\dot{\psi_{1}}^2+\dot{\theta}^2h_{4}^2]
 & \text{i=4}.
 \\2 R^2[\dot{\psi_{2}}^2+\dot{\theta}^2h_{5}^2]
 & \text{i=5}.
 \\2 R^2[\dot{\psi_{3}}^2+\dot{\theta}^2h_{6}^2]
 & \text{i=6}.
 \\2R^2[(\dot{\psi_{1}}+\dot{\psi_{2}})^2+\dot{\theta}^2h_{7}^2]
 & \text{i=7}.
 \\2R^2[(\dot{\psi_{1}}+\dot{\psi_{3}})^2+\dot{\theta}^2h_{8}^2]
 & \text{i=8}.
 \\2R^2[(\dot{\psi_{2}}+\dot{\psi_{3}})^2+\dot{\theta}^2h_{9}^2]
 & \text{i=9}.
\\2[(\dot{R}+R\dot{\psi_{1}})^2+R^2\dot{\theta}^2h_{10}^2]
 & \text{i=10}.
 \\2[(\dot{R}+R\dot{\psi_{2}})^2+R^2\dot{\theta}^2h_{11}^2]
 & \text{i=11}.
 \\2[(\dot{R}+R\dot{\psi_{3}})^2+R^2\dot{\theta}^2 h_{12}^2]
 & \text{i=12}.
\end{cases}
\end{equation}
After a straightforward calculation one obtains:
 \begin{eqnarray} \label{eis0}
&&\varepsilon_{o}=\dot{R}^2+(\nabla R)^2+R^2[\dot{\theta}^2+(\nabla\theta)^2]+V(R),\\ \label{eis1}
&&\varepsilon_{1}=B{\cal K}_{1}^2R^2[5\dot{\theta}^2+ (\nabla\theta)^2+2],\\ \label{eis2}&&
\varepsilon_{2}=B{\cal K}_{2}^2[5R^2h_{3}^2\dot{\theta}^2+5\dot{R}^2+R^2h_{3}^2(\nabla\theta)^2+(\nabla R)^2+2(h_{2}+1)^2R^2],\\ \label{eis3}&&
\varepsilon_{3}=B{\cal K}_{3}^2[5(R h_{3}\dot{\theta}+\dot{R})^2+(R h_{3}\nabla\theta+\nabla R)^2+2(h_{3}+1)^2R^2],\\ \label{eis4}&&
\varepsilon_{4}=B{\cal K}_{4}^2R^2[h_{4}^2(5\dot{\theta}^2+(\nabla \theta)^2) +5\dot{\psi_{1}}^2+(\nabla\psi_{1})^2+2(h_{4}+1)^2],\\
\label{eis5}&&
\varepsilon_{5}=B{\cal K}_{5}^2R^2[h_{5}^2(5\dot{\theta}^2+(\nabla \theta)^2) +5\dot{\psi_{2}}^2+(\nabla\psi_{1})^2+2(h_{5}+1)^2],\\
\label{eis6}&&
\varepsilon_{6}=B{\cal K}_{6}^2R^2[h_{6}^2(5\dot{\theta}^2+(\nabla \theta)^2) +5\dot{\psi_{3}}^2+(\nabla\psi_{3})^2+2(h_{6}+1)^2],\\ \label{eis7}&&
\varepsilon_{7}=B{\cal K}_{7}^2R^2[h_{7}^2(5\dot{\theta}^2+(\nabla \theta)^2)+5(\dot{\psi_{1}}+\dot{\psi_{2}})^2+(\nabla\psi_{1}+\nabla\psi_{2})^2 +2(h_{7}+1)^2],\\ \label{eis8}&&
\varepsilon_{8}=B{\cal K}_{8}^2R^2[h_{8}^2(5\dot{\theta}^2+(\nabla \theta)^2)+5(\dot{\psi_{1}}+\dot{\psi_{3}})^2+(\nabla\psi_{1}+\nabla\psi_{3})^2 +2(h_{8}+1)^2],\\ \label{eis9}&&
\varepsilon_{9}=B{\cal K}_{9}^2R^2[h_{9}^2(5\dot{\theta}^2+(\nabla \theta)^2)+5(\dot{\psi_{2}}+\dot{\psi_{3}})^2+(\nabla\psi_{2}+\nabla\psi_{3})^2 +2(h_{9}+1)^2],
\\ \label{eis10}
&& \varepsilon_{10}=B{\cal K}_{10}^2[R^2h_{10}^2 (5\dot{\theta}^2+(\nabla\theta)^2)+5(\dot{R}+R\dot{\psi_{1}})^2+(\nabla R+R\nabla\psi_{1})^2+2R^2(h_{10}+1)^2],~~~~~~~~~\\ \label{eis11}&&
\varepsilon_{11}=B{\cal K}_{11}^2[R^2h_{11}^2 (5\dot{\theta}^2+(\nabla\theta)^2)+5(\dot{R}+R\dot{\psi_{2}})^2+(\nabla R+R\nabla\psi_{2})^2+2R^2(h_{11}+1)^2],~~~~~~~~~
\\ \label{eis12}&&
\varepsilon_{12}=B{\cal K}_{12}^2[R^2h_{12}^2 (5\dot{\theta}^2+(\nabla\theta)^2)+5(\dot{R}+R\dot{\psi_{3}})^2+(\nabla R+R\nabla\psi_{3})^2+2R^2(h_{12}+1)^2],~~~~~~~~~
\end{eqnarray}
All terms in the above relations are positive definite except (\ref{eis0}). Moreover,  all  brackets $[\cdots]$ in relations (\ref{eis1})-(\ref{eis12}) are multiplied  by one of the  ${\cal K}_{i}^2$'s ($i=1,2,\cdots,12$). Therefore, all $\varepsilon_{i}$'s ($i=1,2,\cdots,12$) are positive definite and are zero simultaneously just  for the non-trivial special solution (\ref{ss}) (and the trivial  vacuum state $R=0$).  For the other solutions, at least one of the  ${\cal K}_{i}$'s is a    nonzero functional, thus at least one of the $\varepsilon_{i}$'s ($i=1,2,\cdots,12$) would be a nonzero positive definite function. Now, if one considers a system with a large value of parameter $B$, then for other solutions, the term  $\sum_{i=1}^{12}\varepsilon_{i}$ would be a large positive definite function which leads to total energies larger than the rest energy of the special solution (\ref{ss}).

More precisely, to confirm that the special solution (\ref{ss}) is energetically stable, it is necessary to examine the energy density (\ref{nnmn}) for any arbitrary small deformations above the background of that when it is at rest.  In general, any arbitrary small  deformed version  of the special solution (\ref{ss})  can be introduced  as follows:
\begin{equation} \label{so1}
R=R_{s}+\delta R \quad \textrm{and} \quad \theta=\theta_{s}+\delta \theta, \quad \psi_{j}=\psi_{js}+\delta \psi_{j}\quad (j=1,2,3),
\end{equation}
where $\delta R$, $\delta \theta$ and $\delta \psi_{j}$  (small variations) are  considered to be any arbitrary  small functions  of space-time. Note that, $R_{s}=\exp{(-r^2/2)}$, $\theta_{s}=\omega_{s}t$ and $\psi_{js}=x^{j}\exp{(-r^2/2)}$ ($j=1,2,3$).
Now, if we insert  (\ref{so1}) into    $\varepsilon_{o}$ and keep it to the first order of $\delta R$ and $\delta \theta$, then it yields
\begin{eqnarray} \label{so3}
&&\varepsilon_{o}=\varepsilon_{os}+\delta\varepsilon_{o}\approx \left[(\nabla R_{s})^2+R_{s}^2\omega_{s}^2+V(R_{s})\right]+\nonumber\\&&
\quad\quad\quad 2\left[\nabla R_{s} \cdot \nabla(\delta R)+R_{s}(\delta R)\omega_{s}^2+
   R_{s}^{2}\omega_{s}(\delta\dot{\theta})+\frac{1}{2}\frac{dV(R_{s})}{dR_{s}}(\delta R)\right].
\end{eqnarray}
Note that, for the non-moving special solution (\ref{ss}), $\dot{R_{s}}=0$, $\nabla\theta_{s}=0$ and $\dot{\theta_{s}}=\omega_{s}=\sqrt{2}$. It is obvious that $\delta\varepsilon_{o}$ is not necessarily a positive definite function.

Now, let do this for the additional terms $\varepsilon_{i}$ ($i=1,2,\cdots,12$). If we insert  a variation like (\ref{so1}) into  $\varepsilon_{i}$ ($i=1,2,\cdots,12$), it yields
\begin{eqnarray} \label{so4}
&&\varepsilon_{i}=\varepsilon_{is}+\delta\varepsilon_{i}=\delta\varepsilon_{i}=B[3(C_{is}+\delta C_{i})({\cal K}_{is}+\delta{\cal K}_{i})^{2}-({\cal K}_{is}+\delta{\cal K}_{i})^{3}]=\nonumber\\&&
B[3(C_{is}+\delta C_{i})(\delta{\cal K}_{i})^{2}-(\delta{\cal K}_{i})^{3}]\approx B[3C_{is}(\delta{\cal K}_{i})^{2}-(\delta{\cal K}_{i})^{3}]\approx[3BC_{is}(\delta{\cal K}_{i})^{2}]>0,
\end{eqnarray}
in which $\varepsilon_{is}=0$, ${\cal K}_{is}=0$ and $C_{is}$ referred  to the special solution (\ref{ss}). Since  $\delta{\cal K}_{i}$ and $\delta C_{i}$  are in the first  order of variations $\delta R$, $\delta \theta$ and $\delta\psi_{j}$ ($j=1,2,3$), hence according to Eq.~(\ref{so4}),  $\delta\varepsilon_{i}$ would   be in the second order of the variations. Therefore, since in general $C_{i}>0$, according to Eq.~(\ref{so4}),
$\delta\varepsilon_{i}=\varepsilon_{i}$ ($i=1,2,\cdots,12$) are always positive definite for small variations (as were  perviously  obtained  from Eqs.~(\ref{eis1})-(\ref{eis12}) generally).

In general, if for any arbitrary small deformations   $\delta R$, $\delta \theta$ and $\delta\psi_{j}$,  the variation of the  energy density $\delta\varepsilon=\delta\varepsilon_{o}+\sum_{i=1}^{12}
\delta\varepsilon_{i}$ to be always positive definite, certainly the energetically  stability of the special solution (\ref{ss}) is guaranteed properly.  Since $\delta\varepsilon_{o}$ is a linear functional   of the first order of variations and $\sum_{i=1}^{12}
\delta\varepsilon_{i}$ is a  linear functional of the second order of variations, this requirement is not confirmed in general.  However, since  $\delta\varepsilon_{i}$'s ($i=1,2,\cdots,12$) contain large number $B$ but $\delta\varepsilon_{o}$ does not, therefore the comparison between  $\sum_{i=1}^{12}
\delta\varepsilon_{i}$, which are always positive definite, and $\delta\varepsilon_{o}$, which is not necessarily positive, needs more considerations. For example, for three cases $B=1$, $B=10^2$ and $B=10^{40}$, it is obvious that   $|\delta R|<B(\delta R)^2$  for the variations with the magnitudes  larger than $|\delta R|>1$, $|\delta R|>10^{-1}$ and $|\delta R|>10^{-20}$, respectively. Exactly  the same argument goes for the comparison between $|\delta\varepsilon_{o}|$ and $\sum_{i=1}^{12}
\delta\varepsilon_{i}$. In other words, for example, consider a system with  $B=10^{40}$, then   the  order of magnitude of  variations $\delta R$, $\delta\theta$ and $\delta\psi_{j}$ for which the special solution (\ref{ss}) is not mathematically a stable object (i.e. the variations for which  $O(|\delta\varepsilon_{o}|)>O(\delta\varepsilon_{i})\approx O(B(\delta{\cal K}_{i})^{2})$),  is approximately less than $10^{-20}$, which is so small that physically can be ignored  in the stability considerations!
For such so small variations, the total rest energy $E_{o}$ may be reduced  with a very small amount equal  to the integration of $\delta\varepsilon_{o}$ over  the whole  space which again is a very small unimportant value. Therefore for a large value of $B$, the  special solution (\ref{ss})  is effectively an energetically   stable object.

Note that, since  scalars ${\cal K}_{i}$'s  (or  $\mathbb{S}_{i}$'s) are twelve  independent functionals  of $R$, $\theta$ and $\psi_{j}$ ($j=1,2,3$), therefore,  for any arbitrary small deformations, at least one of ${\cal K}_{i}$'s  changes and takes non-zero values. Thus, according to Eq.~(\ref{so4}) and since $B$ is considered to be a large number, $\sum_{i=1}^{12}
\delta\varepsilon_{i}$ changes to be a large positive nonzero function  which  leads  to  a  large increase in  the total  energy. Although  $B$ is consider to be a   large number,   but it  does not affect the dominant dynamical equations (\ref{e25}) and (\ref{e252}) and the observable of the special solution (\ref{ss}).

If one considers a system with an extremely  large value of $B$,  the other (stable) configurations of the fields $R$, $\theta$ and $\psi_{j}$ ($j=1,2,3$),   which are not  close to the special solution (\ref{ss}) and the vacuum state $\varphi=\psi_{j}=0$, requires  extreme  energy to be created.  Thus the single  non-trivial  (stable) configuration  of the fields  with the limited energy  just would be the special solution (\ref{ss}). Since there is not infinite energy in the word, hence the other (stable) configuration of the fields never can be possible to be created. In other words, the new extended system just yields the special solution (\ref{ss}) as the quanta of the system classically.

\begin{figure}[ht!]
  \centering
  \includegraphics[width=150mm]{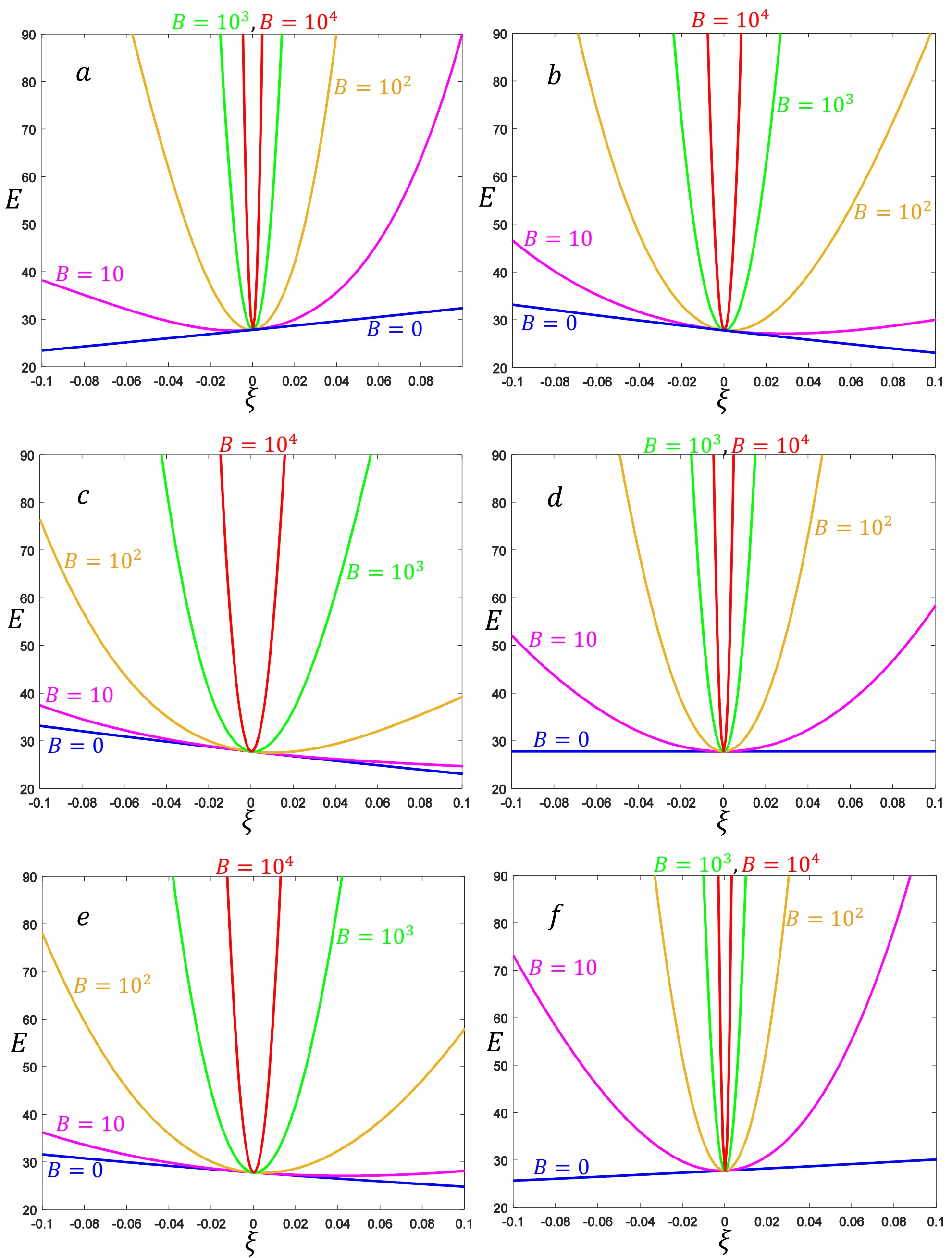}
  \caption{Variations  of  the total rest energy $E$ versus small $\xi$  and different $B$ at $t=0$.  The Figs a-f are related to different variations (\ref{yh1})-(\ref{yh6}), respectively. The case $B=0$  belongs   to  the same original CNKG system (\ref{Lag2}) with the potential (\ref{fp}), and clearly it is not an energetically stable Q-ball solution,  as we expected.  As seen in the Figure,  the larger the values $B$ the greater will be the increase in the total energy for any arbitrary small variation above the background of the special  Q-ball solution (\ref{ss}). Note that, all graphs cross a same point   $(\xi=0,E\approx27.84)$.} \label{1}
\end{figure}

For a better understanding, for example,  we consider   six different arbitrary deformations to show numerically how larger values of parameter $B$ lead to more stability.  Six arbitrary  deformations above the background of the special solution (\ref{ss}) can be  introduced as follows:
\begin{eqnarray} \label{yh1}
&&R=(1+\xi)\exp\left(\frac{-r^2}{2}\right),\quad \theta=\omega_{s}t,\quad \psi_{j}=x^{j}\exp\left(\frac{-r^2}{2}\right), \\&&\label{yh2}
R=\exp\left(\frac{-(r+\xi)^2}{2}\right),\quad \theta=\omega_{s}t,\quad \psi_{j}=x^{j}\exp\left(\frac{-r^2}{2}\right), \\&&\label{yh3}
R=\exp\left(\frac{-(r+\xi)^2}{2}\right),\quad \theta=\omega_{s}t,\quad \psi_{j}=x^{j}\exp\left(\frac{-(r+\xi)^2}{2}\right), \\&&\label{yh4}
R=\exp\left(\frac{-r^2}{2}\right),\quad \theta=\omega_{s}t,\quad \psi_{j}=(1+\xi)x^{j}\exp\left(\frac{-r^2}{2}\right), \\&&\label{yh5}
R=\exp\left(\frac{-(1+\xi)r^2}{2}\right),\quad \theta=\omega_{s}t,\quad \psi_{j}=x^{j}\exp\left(\frac{-r^2}{2}\right), \\&&\label{yh6}
R=\exp\left(\frac{-r^2}{2}\right),\quad \theta=(1+\xi)\omega_{s}t,\quad \psi_{j}=x^{j}\exp\left(\frac{-r^2}{2}\right),
\end{eqnarray}
where $j=1,2,3$ and $\xi$  is a small parameter which  can be considered as an indication of the amount of deformations (variations).  For all  deformed solutions   (\ref{yh1})-(\ref{yh6}), the variation of the total energy versus $\xi$  are shown  in Fig.~\ref{1}~($a$-$f$) respectively. These  figures show that clearly  how the larger values of the parameter $B$ lead
to more stability, i.e. the larger values of $B$ lead to further increase in the total energy versus
$|\xi|$. Note that, the  case $\xi=0$ would be the same non-deformed special solution (\ref{ss}) which its (rest)  energy, according to Eq.~(\ref{fe}) with $l=\lambda=1$, $a=e^{(1)}$ and $\omega_{o}=\sqrt{2} t$,  is  $E_{o}\approx 27.84$. Based  on the Fig.~\ref{1}~($a$-$f$), the case $\xi=0$ would be a minimum   for  the systems with large values of the parameter $B$.  In other words, for  the systems with large values of parameter $B$, the  special solution (\ref{ss}) is  stable against any arbitrary deformation.   The  complementary arguments about these  figures   are the same as those  written in the section 5 of the Ref.~\cite{MM1}.

\section{Summary and conclusion}\label{sec5}

We reviewed some basic properties of the  relativistic $U(1)$-Lagrangian densities which  bear   Q-ball solutions. Especially an  example was introduced in $3+1$ dimensions  which yields   infinite  Gaussian Q-ball solutions. Also, we reviewed all stability criteria  which are used for the Q-ball solutions in the introduction. They are the classical, the fission, the quantum mechanical and the energetically stability  which were explained to the extent necessary. Based on the different stability criteria,  we considered  the stability of the introduced Gaussian Q-ball solutions in detail. Since  none of the Q-balls are essentially energetically stable \cite{MM1},  we add a proper term $F$ to the original standard $U(1)$-Lagrangian density (\ref{Lag2}) to  guarantee the energetically stability of  a special    (Q-ball) solution (\ref{ss}).  Moreover, this proper additional term is constructed deliberately  in such a way whose  role in the dominant dynamical equations and other properties of the special (Q-ball) solution (\ref{ss}) being  ineffective. Briefly, it behaves as a stability catalyzer just for the special solution (\ref{ss}). In order to fulfill  the requested roles by the additional term $F$,   three new catalyzer fields $\psi_{j}$ ($j=1,2,3$) must be included.

The special (Q-ball) solution (\ref{ss}) is a  single solution among the others; meaning that, there is no other solutions with the same properties  of  the special solution (\ref{ss}). In other words, just for the  special solution (\ref{ss}), all complicated dynamical equations (\ref{jkt})-(\ref{jkt2}) and energy density function (\ref{nnmn})  are reduced to the same original versions (\ref{e25}), (\ref{e252}) and (\ref{TE}), respectively.
It was shown that for any arbitrary  small variation above the background of the  special solution (\ref{ss}), the total  energy always increases.   In other words, the special solution (\ref{ss}) is energetically stable.


\end{document}